\documentclass{article}
\usepackage{ijcai16}
\usepackage{amsmath, amsthm, amssymb}
\usepackage{graphicx}

\usepackage{times}

\makeatletter
\def\hlinewd#1{%
  \noalign{\ifnum0=`}\fi\hrule \@height #1 \futurelet
   \reserved@a\@xhline}
\makeatother

\title{Cross-Domain Entity Resolution in Social Media % $^\dag$
}
\author{
W. M. Campbell, Lin Li, C. Dagli, \\
{\bf J. Acevedo-Aviles, K. Geyer,}\\
{\bf J. P. Campbell} \\
Human Language Technology Group \\
MIT Lincoln Laboratory\\
Lexington MA \\
{\small \{wcampbell,lin.li,cdagli,joel,kgeyer,jpc\}@ll.mit.edu}
\And
C. Priebe\\
Department of Applied Math and Statistics \\
Johns Hopkins University\\
Baltimore MD \\
{\small cep@jhu.edu}}

\begin{document}
\maketitle

%
% Footnotes
%
\renewcommand{\thefootnote}{$\dag$} \footnotetext{This work was sponsored
by the Defense Advanced Research Projects Agency under Air Force
Contract FA8721-05-C-0002. Opinions, interpretations, conclusions,
and recommendations are those of the authors and are not necessarily
endorsed by the United States Government.}
% \renewcommand{\thefootnote}{$^*$} \footnotetext{The author is now with MIT Lincoln Laboratory.}

%
% Abstract
%
\begin{abstract}
%
% Abstract
%

The challenge of associating entities across multiple domains is a key
problem in social media understanding.  Successful cross-domain entity
resolution provides integration of information from multiple sites to
create a complete picture of user and community activities,
characteristics, and trends.  In this work, we examine the problem of
entity resolution across Twitter and Instagram using general
techniques.  Our methods fall into three categories: profile,
content, and graph based.  For the profile-based methods, we
consider techniques based on approximate string matching.  For content-based methods, we perform author identification.  Finally, for graph-based methods, we apply novel cross-domain community detection methods
and generate neighborhood-based features.  The three categories of
methods are applied to a large graph of users in Twitter and
Instagram to understand challenges, determine performance, and
understand fusion of multiple methods.  Final results demonstrate an
equal error rate less than $1\%$.

\end{abstract}

%
% Intro
% 
\section{Introduction}\label{sec:intro}
%
% Introduction
%

We consider the problem of associating entities (typically people and
organizations) across multiple social media sites.  Specifically,
given an entity's username {\tt @org12} on Twitter, can we
identify, if it exists, the corresponding username {\tt @org123} for the same
entity on Instagram.  We call this problem {\it cross-domain entity
resolution}.

Strategies for entity resolution are multifold.  In a document data
set, cross-document entity coreference
resolution~\cite{bagga1998entity} seeks to identify text mentions that
correspond to the same entity.  In another task, entity linking, the
goal is to associate entity mentions in text to a knowledge
base~\cite{han2011collective,rao2013entity}.  Entity resolution, also
known as record linkage or de-duplication, in relational data and
databases is addressed
in~\cite{bhattacharya2007collective,christen2012survey}.

The cross-domain entity resolution problem is strongly related to the
above mentioned methods, but is distinct.  In social media, information
about an entity is present in multiple forms as profiles, content, and
graph structure.  Profiles give information about username,
full name, profile pictures, links, etc.  Content shows the topics and
idiolect of a particular entity.  Finally, graph structure relates the
local communications and interests of the entity.  

Prior approaches to the entity resolution problem in social media are
given
in~\cite{bartunov2012joint,goga2013exploiting,iofciu2011identifying,liu2013s,lyzinski2015,malhotra2012studying,peled2013entity,raad2010user,tan2014mapping,zhang2015multiple,zhang2015cosnet}.
Peled, et. al. is the most complete and explores entity resolution
with multiple features, including name, document similarity, and graph
features.  Our approach focuses on more challenging cross-domain
platforms (Twitter and Instagram) at large scale with noisy,
incomplete information. In addition, we use more advanced features in
each category.  We propose multiple normalization methods for profile
matching, including a novel application of the Burrows-Wheeler
transform, and additional token-based string comparison metrics.  On
the graph side, we construct a graph using both usernames and
hashtags.  We then apply a recent cross-domain community detection
algorithm to generate additional graph features~\cite{Li2015}.  Our
extensive experiments on Twitter and Instagram show the best
within-category methods and also demonstrate the fusability of the
different features.

The outline of this paper is as follows.  In Section~\ref{sec:setup},
we outline the basic problem and data setup.  In
Section~\ref{sec:profile}, we describe profile-based methods and our
approximate string matching techniques.  In Section~\ref{sec:content},
we review author identification approaches.  Section~\ref{sec:graph} discusses 
methods to perform joint community detection and the resulting
graph-based features.  Finally, Section~\ref{sec:exp} applies the
methods to large Twitter and Instagram data sets and fuses the
results.

%
% Setup
%
\section{Problem Setup}\label{sec:setup}
%
% Problem Setup
%
\subsection{Entity Resolution in Cross Media}
Our goal is to match entities across different social media platforms.
Information from this process is obtained from three sources.  First,
profile information from the JSON object for a post is used to obtain
attributes of the entity (we use `post' to describe posts on
Instagram and tweets on Twitter).  To create a general approach, we
use only the username and (full) name of the entity as profile
features.  A second source of information is the content of the posts.
Third, we construct a graph using mentions of hashtags and users and
extract graph-based features; see Section~\ref{sec:graph} for more
details.

Our basic approach to the problem is to create comparison scores
between multiple features for entities and then fuse the resulting
scores.  For the graph-based features, we assume the presence of {\it
  seeds}.  Seeds are known or high-confidence entity matches between
different social media platforms.

Multiple problems are encountered in the process of matching.  First,
Twitter and Instagram are noisy  platforms.  Information about a
user is self-reported in many cases and may be misleading or
erroneous.  Additionally, posts between users may not indicate a
relation between individuals; this process creates a noisy graph.  A
second related problem is incomplete information.  Many profiles for
users are unavailable---their accounts are private or have been
removed.  Also, individuals may leave profile fields blank, post
infrequently, or not communicate with others.  These behaviors all
create incomplete information in the proposed features.  A third
challenge, which we do not address in this paper, is sampling. Both
the graph and content features are dependent on the duration and
coverage of the sampling of posts.  Overall, we address these
challenges through feature robustness and fusion.

\subsection{Data Set}
Geotagged Twitter and Instagram data from the Boston area was
collected for our experiments. Twitter data consist of approximately
4.65 million tweets collected from 1/1/2014 to 10/1/2014.  Instagram
data consist of 3.71 million posts (and comments) collected between
12/31/2013 to 12/31/2014. For Instagram, some comments on these posts
extended into 2015.

%
% Profile
%
\section{Profile-Based Methods}\label{sec:profile}
%
% Profile based approaches
%

Profile-based entity resolution relies upon comparing user attributes
and finding measures of similarity.  Depending on the social media
platform, different profile information is available.  For generality,
we focus on usernames and full names.

Multiple observations can be made for these attributes. For
usernames that may not be available across platforms, a common strategy is for users to add a few extra
characters to their username on one platform to obtain a username on
another platform.  As a result, approximate string matching is an appropriate
comparison method.  Also, users may rearrange substrings in a
username; e.g., change {\tt @bobsmith} to {\tt @smithbob}.  For this
case, we use a transform to normalize the
rearrangement---Burrows-Wheeler.  Finally, variations in case, unusual
characters, etc., especially in full names, are addressed with text
normalization.

\subsection{Approximate String Matching Techniques}
Our pipeline for approximate string matching is as follows.  First, we
optionally perform text normalization on the string.  Normalization
converts any unusual UTF-8 characters to a standardized form,
eliminates emojis, and eliminates emoticons. We also remove
nonsentential punctuation, markup, and long repeats (e.g., `cooooool'
to `cool'). Full names are reordered to a standard form (e.g.,
`Smith, Bob' to `Bob Smith').  Second, we optionally lowercase
the entire string.  Third, we compute a normalized similarity between
strings.

For approximate string metrics, we use standard
methods---Damerau-Levenshtein, Levenshtein, Jaro, Jaro-Winkler, and
Jaro-Winkler with soft-TFIDF; see, for
example~\cite{cohen2003comparison}.  To convert Levenshtein and
Damerau-Levenshtein to similarity measures with a $[0,1]$ range, we
apply the transform from~\cite{yujian2007normalized}.  To obtain a
similarity from Jaro and Jaro-Winkler, we use $1$ minus the distance.

\subsection{Burrows-Wheeler Transform}
Users naturally segment and rearrange substrings in usernames to
create new usernames.  Finding matches in this case is an interesting
and challenging problem.  A natural solution, which is difficult to
implement, is to segment a username into tokens and then perform
approximate token matching.  For example, a username like
{\tt @bobtsmith} might be segmented to `bob', `t', and `smith' and
then comparisons could be based on the match of these tokens.  This
process requires a trained system to split the character stream; i.e.,
how do we infer where spaces have been removed.

As an alternate to a full tokenization approach, we consider only
limited movement of tokens within a string.  Specifically, only
circular shifts of tokens are allowed. For example, {\tt @bobtsmith}
to {\tt @smithbobt}, but not to {\tt @tbobsmith}.  We use a lossy
version of the Burrows-Wheeler (BW) transform~\cite{burrows1994block}
on the input string.  We {\it do not} add a terminate character to the
input.  We then, as in the standard BW transform, find all circular
shifts of the string. Next, the strings are sorted lexicographically.
The BW transform is the concatenation of the last character of each
string in the sorted list.  For instance, in our example, we obtain the same BW transform
{\tt @hotmsbtib} for both inputs {\tt @smithbobt} and {\tt @tbobsmith}.  Note this
avoids comparing all possible shifts of usernames to find the best
alignment.

%
% Content 
%
\section{Content-Based Methods}\label{sec:content}
%
% Content
%

\newcommand{\counts}{\operatorname{count}}

For content-based entity disambiguation, we use standard methods for
author identification based on idiolect~\cite{WCampbell07b}.  
For each Twitter user $U$, we collect all tweets.  We then perform
text normalization to eliminate unusual UTF-8 characters (emojis,
emoticons), links, and any repeated characters.  Then, counts of the
words and hashtags are found, $\counts(w_i|U)$, where $w_i$ is the
$i$th word in the dictionary of possible words (including hashtags).

As in standard text classification, the counts are converted to
a vector ${\bf v}$ with weighted probability entries
\begin{equation}
v_i = c_i p(w_i|U)
\end{equation}
where
$
p(w_i|U) = \frac{\counts(w_i|U)}{\sum_V\counts(w_i|V)}.
$
We use a log weighting of
$
c_i = \log\left(\frac{1}{p(w_i|{\rm all})}\right)+1
$,
where $p(w_i|{\rm all})$ is the probability of word $w_i$ across all Twitter
users.

To train author identification models, we use a Support Vector Machine (SVM) one-vs-rest approach; i.e.,
we train each Twitter user $U$ against the remaining Twitter users to
obtain an SVM $f_U(\cdot)$.  For our approach, a linear kernel is
used.  Additionally, models are built only if a minimum number of
words (200, empirically determined) was available for that user.
This minimum word count ensures that the training process is
well-conditioned, but results in many users not having models.

To produce scores, we again find a single vector per user based on the
Instagram posts.  These vectors are scored with the Twitter SVM
models, $f_U(\cdot)$, for all trials.  Note that the role of Twitter
and Instagram for training and testing can be switched, but we use
only one direction for simplicity.

%
% Graph
%
\section{Graph-Based Methods}\label{sec:graph}
%
% Graph-based approaches
%
The graph-based approach extracts user features based on the graph
structure and computes a similarity between the graph features. Examples of graph features are community membership and
(weighted) neighborhood. Before describing the extraction of different graph
features, it is important to first construct graphs that capture the
richness of both Twitter and Instagram data.
%
% Graph construction
%
\subsection{Content + Context Graph Construction}\label{sec:graph_construct}
%
% Graph Construction
%

Graph construction is performed by designating both users (e.g.,
{\tt @twitter}) and hashtags (e.g., {\tt \#fashion}) as vertices in the graph.
For Twitter, edge types correspond to multiple
categories---user-to-user tweets, user mentions of users or hashtags,
retweets, and co-occurrence of hashtags or users.  For Instagram, edge
types correspond to---user comments on user posts, user mentions of
users, user mentions of hashtags, and co-occurrence of users or
hashtags.  For both Twitter and Instagram, the count of occurrence of
various edge types is saved in the graph.  For final analysis,
counts are summed across edge types, resulting in a weighted
undirected graph.  
% More details may be found in~\cite{campbell-baseman-greenfield:2014:SocialNLP}.

%
% Cross-media community detection
%
\subsection{Cross-Domain Community Detection}
%
% Community
%
Using the technique described above, we can construct a Twitter graph
$G_{\rm twitter}$ and an Instagram graph $G_{\rm inst}$. To extract
community features, we perform cross-media community detection to
identify communities simultaneously across Twitter and Instagram
graphs. The key to achieving this is to align the graphs using {\it
seeds}. Seeds are known vertex matches across graphs (e.g., one
obvious choice of seeds is the common hashtags). We use a random
walk-based approach to align the graphs to form a single
interconnected graph. There are three general strategies: (1) {\it
aggregation} that merges pairs of vertices in the seed set; (2) {\it
linking} that adds links to the seed pairs; and (3) {\it relaxed
random walk} that allows a random walker to switch between graphs with
some probability. Once the graphs are aligned and connected, it is
straightforward to adapt Infomap~\cite{rosvall2008maps} for community
detection. Infomap is a random walk-based algorithm that partitions
the graph by minimizing an information-theoretic based objective
function.

For the experiment, we use the aggregation approach with Infomap for
community detection across Twitter and Instagram graphs. Prior
work~\cite{Li2015} shows that with a sufficient number of seeds,
the aggregation approach is the most faithful to the underlying
community structure. Specifically, we first associate a Markov
transition matrix to the {\it union} of $G_{\rm twitter}$ and $G_{\rm
inst}$. Each element in the Markov matrix represents the probability
of a random walk of length $1$ going from one vertex to the other; the
Markov transition probability is computed by normalizing the edge
weights between a vertex and all of its adjacent vertices. Second, for
each vertex pair in seeds, we merge the two vertices and update the
transition matrix with probability $p=0.5$ that a random walk moves to
the adjacent vertices in $G_{\rm twitter}$ and probability $1-p$ that
a random walk moves to the adjacent vertices in $G_{\rm inst}$. The
resulting aligned and connected graph is denoted as $G_{\rm join}$; it
includes all the vertices from both $G_{\rm twitter}$ and $G_{\rm
inst}$ and the edge weights are given by the Markov transition
matrix. Additionally, we apply Infomap only on the largest connected
component of the aligned graph $G_{\rm join}$; vertices that are in
the largest connected component have a community assignment.

%\begin{itemize}
%\item Describe basic approach via random walks and converting social media tensor to a single graph.  
%\item Refer to infomap paper and describe basic idea.  
%\item Describe best approach to simultaneous community detection.  
%\item Refer to amazing NIPS workshop paper on this topic.
%\end{itemize}
%

%
% Graph features
%
\subsection{Graph-Based Features and Similarity Measures}
%
% Graph feats
%
As hinted earlier, we are interested in extracting two classes of
graph features: community features and neighborhood features. Note that
neighbors of a vertex in a graph are vertices connected by an edge to
the specified vertex; they are also referred to as 1-hop
neighbors. Generally, for a vertex $v$ in a graph, $k$-hop neighbors
are defined as vertices that are reachable from $v$ in {\it exactly}
$k$ hops.

\subsubsection{Community Features and Similarity} 
The basic idea is to be able to represent the similarity in community
membership between users across graphs. First, we perform a
cross-media community detection on Twitter and Instagram graphs. One
simple way to compare community features of two users is to assign a
value `1' to the pair that are in the same community and `0'
otherwise. However, this binary-valued similarity score will likely
cause confusion because it assigns a similarity score `1' to all users
belonging to the same community. To mitigate this problem, we propose
to represent a user's community feature via the community membership
of all its (k-hop) neighbors in its respective graph. For example, the
community feature for a Twitter user $U$ is given by a count vector
${\bf c}(U)$ with entries
%% \begin{equation}
%% c_i = \counts\left(i|\operatorname{neighbors}(U) \in G_{\rm
%% twitter}\right)
%% \end{equation}
\begin{equation}\label{eqn:count}
c_i = \left|\left\{N|\operatorname{comm}(N)=i,~N\in\operatorname{nbr}(U|G_{\rm twitter})\right\}\right|
\end{equation}
where $\operatorname{comm}(\cdot)$ indicates the community assignment
of vertex $N$ and $\operatorname{nbr}(\cdot)$ is the set of $k$-hop
neighbors.

For the experiment, we set $k = 1,2$ and use one of the two methods to
measure the similarity in community feature between users. One is to
compute the dot product of normalized count vectors, i.e.,
$\operatorname{sim}({\bf c}(U_i),{\bf c}(U_j))=\frac{{\bf c}(U_i)^T{\bf c}(U_j)}{\|{\bf c}(U_i)\|_2\|{\bf c}(U_j)\|_2}$. The
other method is to use an SVM; it is similar to the method for author
identification discussed in Section~\ref{sec:content}, but replacing a
dictionary of words with a dictionary of communities.

\subsubsection{Neighborhood Features and Similarity}
Given the aligned and connected graph $G_{\rm join}$ (i.e.,
constructed from joining $G_{\rm twitter}$ and $G_{\rm inst}$ using
seeds), we seek to compute the similarity between two users by
analyzing the proximity of their corresponding vertices in $G_{\rm
join}$. A popular approach is based on vertex
neighborhoods~\cite{Liben-Nowell:2003:LPP:956863.956972}, e.g., common-neighbors approach and its variants. The approach here is similar. However,
instead of simply counting the number of common neighbors, we
represent the neighborhood feature using the transition probability of
the random walk in $G_{\rm join}$. Specifically, the neighborhood
feature of a Twitter user $U$ is given by ${\bf p}(U)$, whose $i^{\rm
th}$ entry $p_i = p(U,U_i)$ represents the probability that a random
walk of a specified length $k$ starting at $U$ ends at the $i^{\rm
th}$ vertex $U_i$ in $G_{\rm join}$. The neighborhood similarity is
given by the normalized dot product of the neighorhood features.

We choose to use $k=1,2$ hops for computing the neighborhood
similarity. Note that for $k = 1$, the edge probability $p(U_1, U_2) =
0$ if $U_1$ and $U_2$ are not connected. Also, isolated vertices are
not considered.

%\begin{itemize}
%\item Describe neighborhood community features
%\item Describe common neighbor feature
%\item Describe variations -- hashtag only, etc.
%\end{itemize}

% 
% Fusion
%
%\section{Fusion}\label{sec:fusion}
%\input{fusion}

%
% Experimental Results
%
\section{Experimental Results}\label{sec:exp}
%
% Experimental Results
%

\subsection{Setup}\label{sec:exp_setup}
We use the Twitter and Instagram data sets in Section~\ref{sec:setup}
for our experiments.  For both data sets, profiles are extracted from
the posts, and graphs are constructed using the method in
Section~\ref{sec:graph_construct}.  This processing results in 141.7K
Twitter profiles and 925K Instagram profiles.  The Twitter graph has
860.8K vertices (280.5K hashtags, 580.3K users) and 2.56M edges.  The
Instagram graph has 1.667M vertices (533.7K hashtags, 1.134M users) and
9.706M edges.  Note that not all users have profile information
because a user may be mentioned, but no post is observed from that
user.

We construct a common set of trials to test all systems.  The trials
consist of user pairs $(u_t,u_i)$ where $u_t$ is a user from Twitter
and $u_i$ is a user from Instagram.  True trials (i.e., same entity
trials) are constructed with two strategies: (1) self-reported
Twitter-Instagram links in user profiles and (2) links in tweets to
Instagram to associate users across the different sites.  False trials
are constructed by taking a true trial $(u_t,u_i)$ and randomly
sampling Instagram users $v_i\neq u_i$ to produce trials $(u_t,v_i)$.
Note that only users with one known account on both Twitter and
Instagram are used; users with multiple accounts are discarded.

The resulting number of trials is 25,548 true trials and 255,480 false
trials.  We use five-fold cross validation (train/test) to score all systems.
Results are reported by pooling all test scores across all five splits
and evaluating the results.  We report results for both all trials, which
include all of the trials above, and the non-trivial (NT) trials.
Non-trivial trials are trials where the usernames are not an exact
string match.

\subsection{Evaluation}\label{sec:eval}
% Removed Martin Ref to DETs
We use two methods for evaluation in the context of a detection problem. First, we look at miss $P_m$ and false
alarm $P_{\rm fa}$ for a threshold $T$.  Each trial results in a
score.  If that score belongs to a true trial (i.e., should be an
actual match) and is below $T$, this results in a miss.  If the trial
is a false trial (i.e., not an actual match) and the score is above
the threshold, then a false alarm occurs.  Sweeping the threshold
where $P_m$ equals $P_{\rm fa}$ gives our first method of evaluation,
the equal error rate (EER).  Our second evaluation method is to show
detection error tradeoff (DET) curves.  This allows a
user to view performance at multiple operating points; e.g., low
false alarm probability.

Two issues are addressed in evaluation.  First, we use miss and false
alarm rate instead of precision and recall so that our evaluation is
not dependent on the priors for true and false trials.  Second, we
note when missing trials occur because of missing content or data.
The main impact of missing trials is that cross-feature performance is
difficult to compare exactly.  We address this by comparing only per
feature performance and fusion performance.  The relative merit of individual
features can be approximately inferred from their impact on fusion.

\subsection{Profile and Content-Based Features}
%
% Profile system selection
%

We use the pipeline of four stages for comparing profile usernames
and full names as described in Section~\ref{sec:profile}: {\bf norm}, string normalization;
{\bf lower}, conversion to lower case;
{\bf bw}, lossy Burrows-Wheeler transform (non-invertible);
and {\bf jw/nl/ndl}, the approximate match method--Jaro-Winkler,
  normalized Levenshtein, and normalized Damerau-Levenshtein.
%% \begin{itemize}
%% \setlength{\itemsep}{1pt}
%% \setlength{\parskip}{0pt}
%% \setlength{\parsep}{0pt}
%% \item {\bf norm} : string normalization
%% \item {\bf lower} : conversion to lower case
%% \item {\bf bw} : Lossy Burrows-Wheeler transform (non-invertible)
%% \item {\bf jw/nl/ndl} : the approximate match method--Jaro-Winkler,
%%   normalized Levenshtein, and normalized Damerau-Levenshtein
%% \end{itemize}
A `no' in front of the step indicates it is omitted.  All trials
specified in Section~\ref{sec:setup} are scored for both profile
features.

\begin{figure}[tb!]
\centering
\vspace{-0.15in}
\includegraphics[width=0.8\linewidth]{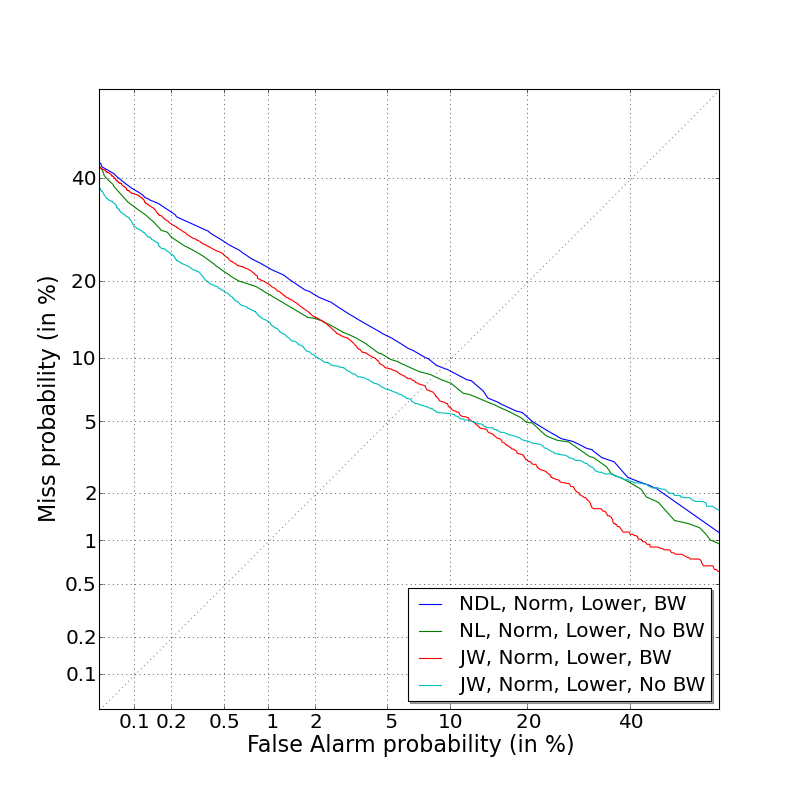}
\vspace{-0.2in}
\caption{DET curve comparing the performance of multiple approximate string matching
  methods on the profile username.  Note that the curves are for {\it
    non-trivial} trials.}\label{fig:username}
\end{figure}

A plot comparing a representative set of systems for comparing
usernames is shown in Figure~\ref{fig:username}.  We make several
observations.  First, only minor performance differences are observed
between Jaro and Jaro-Winkler.  Also, the same is true for Levenshtein
and Damerau-Levenshtein.  Second, normalization and lower case
conversion are not useful (as expected).  Third, the biggest
performance impact is the use of the Burrows-Wheeler type transform.
We observe across many of the cross-validation splits that the BW
transform crosses over with the no BW case in the DET curve.
Sometimes this cross-over is after the EER (as shown) or before.  This
demonstrates that the usefulness of the BW transform depends on the
operating point.

\begin{figure}[tb!]
\centering
\vspace{-0.15in}
\includegraphics[width=0.8\linewidth]{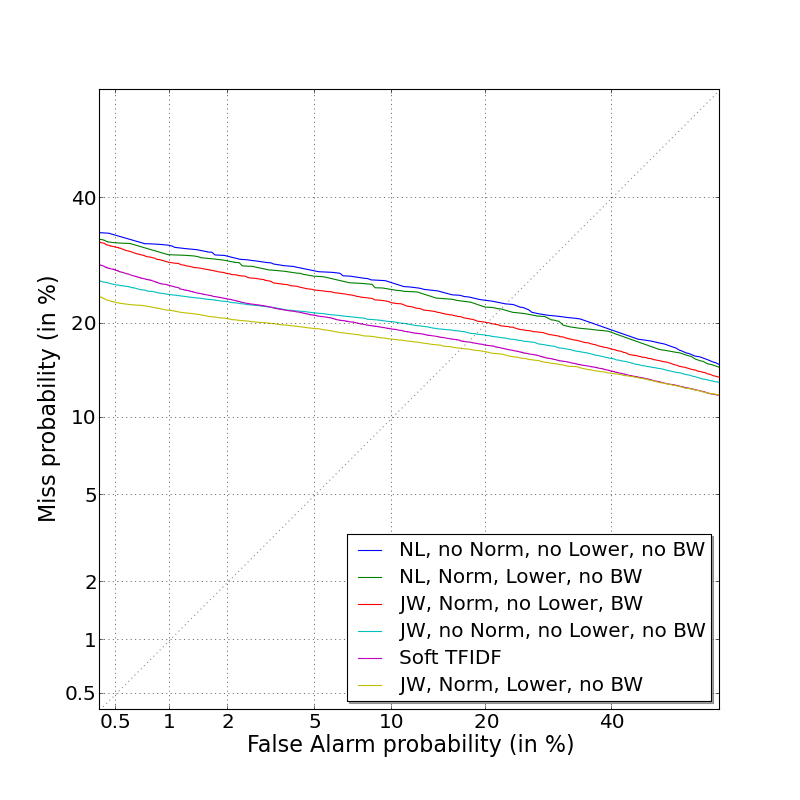}
\vspace{-0.2in}
\caption{DET curve comparing the performance of multiple approximate string matching
  methods on the profile full name.  Note that the curves are for {\it
    non-trivial} trials.}\label{fig:fullname}
\vspace{-0.1in}
\end{figure}

A plot comparing a representative set of systems for full name is
shown in Figure~\ref{fig:fullname}.  The performance is quite distinct
from the username case.  The best performing system uses Jaro-Winkler
and normalizing transforms.  Removing these steps substantially
impacts performance.  In addition, the BW step substantially
lowers performance.  This property might be expected since name
variation is not easily captured as a circular string rotation.
Finally, as before, normalized Levenshtein and Damerau-Levenshtein do
not perform as well as the Jaro-type comparisons.

As a summary of performance at EER, see the results for profile and
content features in Table~\ref{tab:profile}. As a convention in all tables, we bold the metric per feature of the best performing system. In the table, for the full
name feature, we show all of the variations of the pipeline for
Jaro-Winkler approximate string comparison.  Results for the other cases
are comparable.  For both username and full name, Jaro-Winkler with
normalization and lowercase with no Burrows-Wheeler are best at EER.
The table also highlights performance for the content features and the
token-based soft-TFIDF method.  
\begin{table}[tb!]
\centering
\caption{Comparison of EERs for various profile and content features and systems.}\label{tab:profile}
\smallskip\small
\begin{tabular}{|l l l l|}
\hline
{\bf Feature} & {\bf System} & {\bf EER} & {\bf EER } \\ 
              &              & {\bf  All (\%)} & {\bf NT (\%)} \\
\hlinewd{1pt}
Full & jw, norm, lower, bw & 18.73 & 16.94 \\
Full & jw, norm, lower, nobw & {\bf 16.90} & {\bf 14.97} \\ 
Full & jw, norm, nolower, bw & 20.13 & 18.69 \\ 
Full & jw, norm, nolower, nobw & 18.41 & 16.86 \\ 
Full & jw, nonorm, lower, bw & 19.07 & 17.39 \\ 
Full & jw, nonorm, lower, nobw & 17.26 & 15.36 \\ 
Full & jw, nonorm, nolower, bw & 20.32 & 18.95 \\ 
Full & jw, nonorm, nolower, nobw & 18.71 & 17.13 \\
% Full & ndl, norm, lower, bw & 23.27 & 21.62 \\ \hline
% Full & ndl, norm, lower, nobw & 21.95 & 20.19 \\ \hline
% Full & ndl, norm, nolower, bw & 24.26 & 22.92 \\ \hline
% Full & ndl, norm, nolower, nobw & 23.03 & 21.40 \\ \hline
% Full & ndl, nonorm, lower, bw & 22.97 & 21.29 \\ \hline
Full & ndl, nonorm, lower, nobw & 21.61 & 20.04 \\ 
% Full & ndl, nonorm, nolower, bw & 24.22 & 22.96 \\ \hline
% Full & ndl, nonorm, nolower, nobw & 22.70 & 21.35 \\ \hline
% Full & nl, norm, lower, bw & 23.32 & 21.64 \\ \hline
% Full & nl, norm, lower, nobw & 21.98 & 20.23 \\ \hline
% Full & nl, norm, nolower, bw & 24.33 & 23.02 \\ \hline
% Full & nl, norm, nolower, nobw & 23.04 & 21.42 \\ \hline
% Full & nl, nonorm, lower, bw & 23.02 & 21.33 \\ \hline
Full & nl, nonorm, lower, nobw & 21.64 & 20.08 \\ 
% Full & nl, nonorm, nolower, bw & 24.27 & 23.04 \\ \hline
% Full & nl, nonorm, nolower, nobw & 22.71 & 21.38 \\ \hline
Full & soft-tfidf                & 17.69 & 16.24 \\ \hline
User & jw, nonorm, lower, bw & 2.64 & 7.54 \\ 
User & jw, nonorm, lower, nobw & {\bf 2.03} & {\bf 6.45} \\ 
User & ndl, nonorm, lower, bw & 3.17 & 9.06 \\ 
User & ndl, nonorm, lower, nobw & 2.65 & 8.46 \\ 
User & nl, nonorm, lower, bw & 3.14 & 9.24 \\ 
User & nl, nonorm, lower, nobw & 2.66 & 8.52 \\ \hline
Content* & SVM & {\bf 26.56} & {\bf 18.66} \\ \hline
\end{tabular}
\flushleft{\small $^*$subset of all trials because of limited content}
\vspace{-0.1in}
\end{table}

\subsection{Graph-Based Features}
%
% Graph features performance
% 
%
Graph features include community features and neighborhood features (NBR). We
use two methods for computing the community similarity: dot product of
normalized feature vectors (DP) and SVM.  For seed selection, we look
at two different choices of seeds to align Twitter and Instagram
graphs. One is to use common hashtags, for example, {\tt \#Boston} is
present in both graphs and, thus, vertices associated with {\tt \#Boston} can
be used as seeds. The performance of the two graph features is
summarized in Table \ref{table:graph_ht}, showing both the EER for all
trials and EER for the non-trivial trials. Note that for community
feature using $1$-hop neighbors, the SVM method gives better results,
while DP gives better results for community feature using $2$-hop
neighbors.

The other choice of seeds is to use both common hashtags and users
with the same usernames across graphs.  Table \ref{table:graph_both}
shows the performance of the two graph features. In particular, DP is
better than SVM at EER (all trails) for both $1$-hop and $2$-hop
community feature. Note that the EER for all trials is much less than
the EER for non-trivial trials. Because seeds are merged in the
aligned graph, for users with the same username, their neighborhood
features are the same and their community features tend to be similar.

\begin{table}[tb!]
\centering
\caption{Summary of EERs for various community and neighborhood features and systems. Common hashtags are used as seeds.}\label{table:graph_ht}
\smallskip\small
\begin{tabular}{|l l l l|} 
\hline
{\bf Feature} & {\bf System} & {\bf EER ALL (\%)} & {\bf EER NT (\%)}  \\
\hlinewd{1pt}
Comm 1-hop    & DP & 50.13* & 30.22*    \\
Comm 1-hop    & SVM & {\bf 30.47} & {\bf 23.56}     \\
\hline
Comm 2-hop    & DP & {\bf 39.64} & {\bf 30.42}    \\
Comm 2-hop    & SVM & 46.95 & 45.40    \\
\hline
NBR 1-hop & DP & 60.04* & 39.57* \\  	 
\hline
NBR 2-hop & DP & 42.64 & 34.83 \\
\hline
\end{tabular}
\flushleft{\small $^*$interpolated values} \vspace{-0.1in}
\end{table}

\begin{table}[tb!]
\centering
\caption{Summary of EERs for various community and neighborhood features and systems. Common hashtags and users with exact usernames are used as seeds.}\label{table:graph_both}
\smallskip\small
\begin{tabular}{|l l l l|} 
\hline
{\bf Feature} & {\bf System} & {\bf EER ALL (\%)} & {\bf EER NT (\%)}  \\
\hlinewd{1pt}
Comm 1-hop    & DP & {\bf 6.65} & 20.90*    \\
Comm 1-hop    & SVM & 8.33 & {\bf 17.16}     \\
\hline
Comm 2-hop    & DP & {\bf 36.46} & {\bf 29.11}    \\
Comm 2-hop    & SVM & 43.75 & 44.92    \\
\hline
NBR 1-hop & DP & 6.60 & 29.38* \\  	
\hline
NBR 2-hop & DP & 9.38 & 28.47 \\
\hline
\end{tabular}
\flushleft{\small $^*$interpolated values}
\end{table}

\subsection{Fusion}
%
% Fusion
%
%\begin{itemize}
%\item Fusion results
%\item Overall fusion results and curve (table + figure)
%\item Consider subsets -- e.g., best 2, network only, profile only
%\item Tradeoffs with different approaches
%\item Imputation versus multiple fusers
%\end{itemize}

% removed Cox ref, ~\cite{cox58reg} & ~\cite{Breiman:2001:RF:570181.570182}
The goal for fusion is to combine the similarity scores from different
features to obtain a better entity disambiguation system. The models
we train are logistic regression and random forest. The former builds
a linear model that predicts the probability of an outcome using the
logistic distribution function; the latter is an ensemble learning
method that constructs a number of decision trees and averages the
probabilities of an outcome from these decision trees.

One issue for training the fusion models is the problem of missing
data. For example, the content features may be missing for some users
due to an insufficient number of posted words. Similarly, community
(or neighborhood) features may be missing because users are not in the
largest connected component (or they are isolates) in the aligned
graph. One simple strategy to address this issue is to train a
separate model for all combinations of present features using the
training data, and then to test by breaking up the testing set
according to the present features and using the respective model to
compute the fusion score.

\begin{figure}[tb!]
\centering
\vspace{0.05in}
\includegraphics[width=0.8\linewidth]{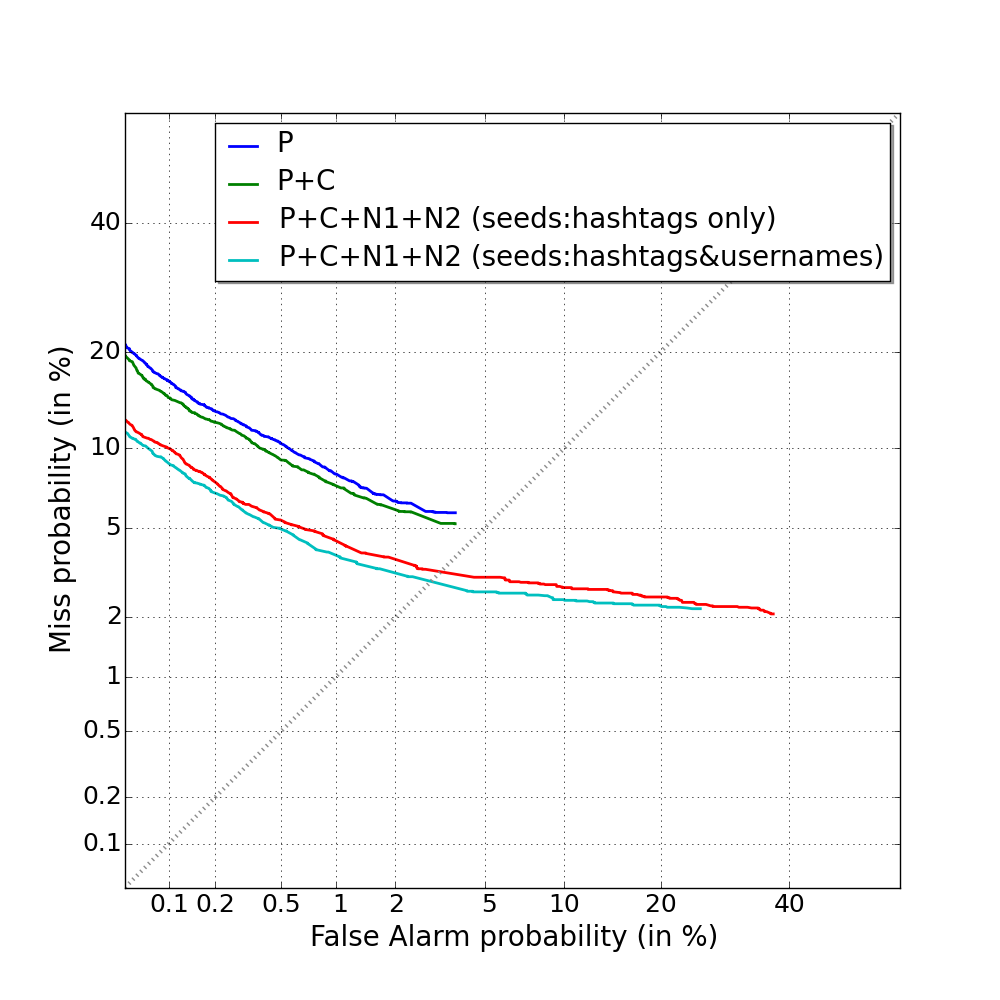}
\vspace{-0.2in}
\caption{DET curve comparing the performances of fusing selected features using the random forest model. Note that the curves are for non-trivial trials.}\label{fig:fusion_RandF}
\end{figure}

Results for fusion are summarized in Tables~\ref{table:fusion_ht}
and~\ref{table:fusion_both}. Features used in the fusion are greedily
selected, i.e., the best system over all trials per feature type. The
variable `P' denotes the fusion of profile features (for the purposes
of this paper, username and full name), `C' denotes content, `N1' is
the fusion of $1$-hop community and $1$-hop neighborhood features, and
`N2' is the fusion of $2$-hop community and $2$-hop neighborhood
features. Figure~\ref{fig:fusion_RandF} shows the performances of
fusing selected features using the random forest model. Observe that
the fusion of profile, content and graph features significantly
improves the performance as compared to the performance of individual
features. Also, the addition of graph features to the fusion model
significantly lowers the miss probability, particularly at the low
false alarm probability. Another observation is that seeding using
common hashtags fuses as well as seeding using both common hashtags
and users with the same usernames.

%\begin{figure}[tb]
%\centering
%\includegraphics[width=0.9\linewidth]{figures/det_fusion_Logit.png}
%\vspace{-0.2in}
%\caption{DET curve comparing the performances of fusion of selected features using logistic regression. Note that the curves are for non-trivial trials.}\label{fig:fusion_Logit}
%\end{figure}

\begin{table}[tb!]
\centering
\caption{Summary of fusion results. Common hashtags are used as seeds.}\label{table:fusion_ht}
\smallskip\small
\begin{tabular}{|l l l l|} 
\hline
{\bf Fusion} & {\bf Model} & {\bf EER ALL (\%)} & {\bf EER NT (\%)}  \\
\hlinewd{1pt}
P   &Logit&  2.07&   6.07  \\
P   &RandF&  {1.54} & {5.74}*    \\
\hline
P+C   &Logit& 2.01 & 5.74   \\
P+C   &RandF& {1.47} & {5.18}*    \\
\hlinewd{0.7pt}
N1    &Logit & {26.74} &  {25.31}  \\
N1    &RandF & 28.44  & 29.44   \\
\hline
N1+N2    &Logit & {26.97}&  26.76  \\
N1+N2    &RandF & 27.43 & {25.91}    \\
\hline
P+N1  &Logit &  1.64&   4.82 \\
P+N1  &RandF & {1.16}  & {3.79}   \\
\hline
P+C+N1 &Logit& 1.64 & 4.87\\  	 
P+C+N1 &RandF& {1.13}& {3.90}* \\  
\hline
P+N1+N2  &Logit &  1.68&   4.85  \\
P+N1+N2  &RandF & {1.01} &  {3.18}  \\	 
\hline
P+C+N1+N2 &Logit& 1.68 & 4.92 \\
P+C+N1+N2 &RandF& {\bf 1.00} & {\bf 3.32} \\
\hline
\end{tabular}
\flushleft{\small $^*$interpolated values}
\vspace{-0.15in}
\end{table}

\begin{table}[tb!]
\centering
\caption{Summary of fusion results. Common hashtags and users with exact usernames are used as seeds.}\label{table:fusion_both}
\smallskip\small
\begin{tabular}{|l l l l|} 
\hline
{\bf Fusion} & {\bf Model} & {\bf EER ALL} (\%) & {\bf EER NT (\%)}  \\
\hlinewd{1pt}
N1    &Logit & {6.54} &  {19.93}  \\
N1    &RandF & 6.87 & 23.11   \\
\hline
N1+N2    &Logit &6.49 & {19.76}   \\
N1+N2    &RandF & {6.44} &  20.43  \\
\hline
P+N1  &Logit & 1.47& 4.56   \\
P+N1  &RandF & {1.08} & {3.43}   \\
\hline
P+C+N1 &Logit& 1.53& 4.66 \\  	 
P+C+N1 &RandF& {1.03}& {3.30}*\\  	
\hline
P+N1+N2  &Logit & 1.44 &  4.44  \\
P+N1+N2  &RandF & {0.94}  & {3.18}   \\ 
\hline
P+C+N1+N2 &Logit&1.48  & 4.56 \\
P+C+N1+N2 &RandF& {\bf 0.89} & {\bf 3.08}  \\
\hline
\end{tabular}
\flushleft{\small $^*$interpolated values}
\vspace{-0.2in}
\end{table}

% \subsection{Analysis}
% \input{exp_analyze.tex}

%
% Experimental Results
%
\section{Conclusions}
%
% Conclusions
%
In this paper, we demonstrated profile-, content-, and
graph-based features for cross-domain entity resolution.  Novel methods for
both profile and graph based features were presented.  Fusion of these
feature types showed that profile and graph features
worked best in combination.  Excellent performance was achieved for
non-trivial entity resolution. We demonstrated our methods on Twitter and Instagram, and we expect our approach to generalize for other social media platforms, as well. 

Future work includes the use of additional features, such as
characteristics of profile images (e.g., properties and hashes),
patterns of life (e.g., geographic tracks over time and activity over
time), and stylometrics (e.g., emojis, emoticons, and unconventional
writing).

%% Additional future work includes the use of alternative
%% fusion methods, application of our cross-domain entity resolution
%% methods to non-English social media platforms, and using hashing to
%% improve the scalability of our methods. Another area of future work is
%% to pursue the complementary challenge of entity resolution within a
%% given social media platform to discover multiple accounts belonging to
%% the same user.

% 
% Ackowledgements ?
%
% \small
\section*{Acknowledgements}
The authors acknowledge inspiration and early
collaboration with Daniel Sussman (Harvard University),
Vince Lyzinski (JHU HLTCOE), and Jason Matterer
(JHU, MIT LL).

% The file named.bst is a bibliography style file for BibTeX 0.99c
\clearpage
\bibliographystyle{named}
\bibliography{entity,graph,recog}

\end{document}